\documentclass[pra,showpacs]{revtex4}

\usepackage{amsmath}
\usepackage{amsfonts}
\usepackage{amssymb}
\usepackage{array}
\usepackage{floatflt}
\usepackage{graphicx}
\DeclareMathOperator{\tr}{Tr}
\newcommand*{\eq}[1]{Eq.~(\ref{#1})}
\newcommand*{\eqs}[2]{Eqs.~(\ref{#1})--(\ref{#2})}

\begin{document}
\title{Evolution of spin entanglement and an entanglement witness in multiple-quantum NMR experiments}

\author{E.~B.~Fel'dman, A.~N.~Pyrkov\footnote{Email address:
pyrkov@icp.ac.ru}} \affiliation{Institute of Problems of Chemical
Physics of Russian Academy of Sciences, Chernogolovka, Moscow
Region, Russia, 142432}

\date{\today}

\begin{abstract}We investigate the evolution of entanglement in
multiple--quantum (MQ) NMR experiments in crystals with pairs of
close nuclear spins-1/2. The initial thermodynamic equilibrium state
of the system in a strong external magnetic field evolves under the
non-secular part of the dipolar Hamiltonian. As a result, MQ
coherences of the zeroth and plus/minus second orders appear. A
simple condition for the emergence of entanglement is obtained. We
show that the measure of the spin pair entanglement, concurrence,
coincides qualitatively with the intensity of MQ coherences of the
plus/minus second order and hence the entanglement can be studied
with MQ NMR methods. We introduce an Entanglement Witness using MQ
NMR coherences of the plus/minus second order.
\end{abstract}


\maketitle

\section{Introduction}

Entanglement~\cite{chuang} is the key concept in Quantum Information
Theory. It has played a crucial role in experiments on quantum
computing and quantum teleportation. This resulted in intensive
interest to the physics of entanglement from both theorists and
experimentalists~\cite{bvsw,amico,ghosh,souza}.

Entanglement is detected with the help of a so-called Entanglement
Witness (EW). By definition, EW is an observable which has a
positive expectation value for separable states and negative for
some entangled states~\cite{horodeski}. In particular, internal
energy~\cite{wang} and magnetic susceptibility~\cite{wiesniak} were
used as EW in some cases. In this paper we propose a new type of an
Entanglement Witness, the intensity of multiple quantum coherences
in spin systems. This quantity is accessible in NMR experiments and
thus opens a new approach to probing entanglement with highly
advanced NMR techniques.

In the present work we focus on the simplest relevant system, a pair
of spins $s=1/2$ coupled by the dipole-dipole interaction in the
conditions of the multiple-quantum (MQ) NMR experiment~\cite{baum}.
Here the initial thermodynamic density matrix describing the
interaction of the spins with the strong external magnetic field is
subjected to the irradiation by the specially tailored sequence of
resonance rf-pulses. The anisotropic dipolar Hamiltonian oscillates
rapidly when the period of the sequence is less than the inverse
dipolar frequency. The spin dynamics of the system is described by
the averaged Hamiltonian which is responsible for the emergence of
the MQ coherences of the zeroth and plus/minus second
orders~\cite{lacelle}. It is evident that the initial state of the
system is separable. However we show with the Wootters
criterion~\cite{wootters} that the entangled state emerges when the
intensity of the MQ coherence of order 2 (-2) exceeds the exactly
calculated threshold depending on the external magnetic field and
the temperature. Thus the intensity of the MQ coherence of the
second order, which is the observable in MQ NMR experiments, serves
as EW for spin systems.

\section{MQ dynamics of a dipolar coupled spin pair at low temperatures}

We consider a two-spin system in a strong external magnetic field
$\vec{H}_0$. The thermodynamic equilibrium density matrix, $\rho_0,$
of the system is
\begin{equation}
\label{rho0} \rho_0=\frac{\exp(\frac{\hbar \omega_0}{kT}I_z)}{Z}
\end{equation}
where $\omega_0=\gamma H_0$ ($\gamma$ is the gyromagnetic ratio),
$T$ is the temperature, $I_\alpha=I_{1\alpha}+I_{2\alpha}$, and
$I_{j\alpha} (j=1,2; \alpha=x,y,z)$ is the projection of the angular
spin momentum operator of spin $j$ on the axis $\alpha$, and $Z$ is
the partition function.

The MQ NMR experiment consists of four distinct periods of time:
preparation, evolution, mixing, and detection~\cite{baum}. MQ
coherences are created by the multipulse sequence consisting of
eight-pulse cycles on the preparation period~\cite{baum}. In the
rotating reference frame~\cite{goldman}, the average Hamiltonian,
$H_{MQ},$ for the two-spin system describing the MQ dynamics at the
preparation period can be written as~\cite{baum}
\begin{equation}
\label{HMQ} H_{MQ}=b\left( I_1^{+}I_2^{+}+I_1^{-}I_2^{-}\right)
\end{equation}
where $b=(\gamma^2 \hbar/\{2r_{12}^3\})(1-3\cos^2\theta_{12})$ is
the coupling constant between spins 1 and 2, $r_{12}$ is the
distance between spins 1 and 2, and $\theta_{12}$ is the angle
between the internuclear vector $\vec{r}_{12}$ and the external
magnetic field $\vec{H}_0;$ $I_j^+$ and $I_j^-$ $(j=1,2)$ are the
rasing and lowering operators of spin $j.$

The two-spin Hamiltonian $H_{MQ}$ can be diagonalized with the
transformation (in the standard basis $\{|00\rangle, |01\rangle,
|10\rangle, |11\rangle\}$)
\begin{equation}
\label{transform} U=\begin{pmatrix} 0&0&\frac1{\sqrt2}&\frac1{\sqrt2}\\
1&0&0&0\\
0&1&0&0\\
0&0&\frac1{\sqrt2}&-\frac1{\sqrt2}
\end{pmatrix},
\end{equation}
and the density matrix, $\rho(\tau),$ at the end of the preparation
period is
\begin{widetext}
\begin{equation}
\label{rhot} \rho(\tau)=e^{-i H_{MQ}\tau}\rho_0e^{i
H_{MQ}\tau}=\frac1{2(1+\cosh\beta)}\begin{pmatrix}
\cosh\beta+\cos(2b\tau)\sinh\beta&0&0&i\sin(2b\tau)\sinh\beta\\
0&1&0&0\\
0&0&1&0\\
-i\sin(2b\tau)\sinh\beta&0&0&\cosh\beta-\cos(2b\tau)\sinh\beta
\end{pmatrix}
\end{equation}
\end{widetext}
where $\beta=\hbar\omega_0/(kT).$ The diagonal part of the density
matrix of \eq{rhot}, $\rho_{(0)}(\tau),$ is responsible for the MQ
coherence of the zeroth order, and the non-diagonal parts,
$\rho_{(2)}(\tau), \rho_{(-2)}(\tau),$ are responsible for the MQ
coherences of the plus/minus second orders\cite{baum,lacelle}. The
intensities of the MQ coherences of the zeroth, $G_0(\tau),$ and
plus/minus second, $G_{\pm2}(\tau),$ orders are\cite{maximov}
\begin{equation}
G_0(\tau)=\tr\left(\rho_{(0)}(\tau)\rho_{(0)}^{ht}(\tau)\right),\qquad
G_{\pm2}(\tau)=\tr\left(\rho_{(2)}(\tau)\rho_{(-2)}^{ht}(\tau)\right),
\end{equation}
where $\rho_{(0)}^{ht}(\tau)$ is the diagonal part of
\begin{equation}
\label{rhoht} \rho^{ht}(\tau)=e^{-i H_{MQ}\tau}I_z e^{i H_{MQ}\tau}
\end{equation}
and $\rho_{(2)}^{ht}(\tau), \rho_{(-2)}^{ht}(\tau)$ are the
non-diagonal parts of the density matrix $\rho^{ht}(\tau).$ Using
\eqs{rhot}{rhoht} one can find that
\begin{equation}
\label{cohs} G_0(\tau)=\tanh\frac{\beta}2\cos^2(2b\tau),\qquad
G_{\pm2}(\tau)=\frac12\tanh\frac{\beta}2\sin^2(2b\tau).
\end{equation}
It is worth to emphasize that intensities of MQ coherences are
observables in MQ NMR experiments. \eq{cohs} shows that the
intensities of the MQ coherences of the second order, $G_2(\tau),$
and the minus second order, $G_{-2}(\tau),$ are equal. However, in
real experiment, certain errors are present and the experimental
results for $G_2(\tau)$ and $G_{-2}(\tau)$ are not the same. Some of
the errors can be compensated and the accuracy can be improved if
one detects the sum of these coherences\cite{yesinowski}. It is also
worth to notice that the accuracy of the measurement of
$G_2(\tau)+G_{-2}(\tau)$ is higher than for $G_0(\tau)$
\cite{yesinowski}. Below we will use the sum of the MQ coherences of
the plus/minus second order in order to introduce the entanglement
witness.

\section{Concurrence and entanglement witness in MQ NMR experiments}

The initial state of the system determined by \eq{rho0} is
separable. Entanglement appears in the course of the preparation
period of the MQ NMR experiment when the MQ coherence of the second
order has a sufficiently large intensity. In order to estimate the
entanglement quantitatively we apply the Wootters
criterion~\cite{wootters}. According to~\cite{wootters}, one needs
to construct the spin-flip density matrix
\begin{equation}
\label{flip}
\tilde{\rho}(\tau)=(\sigma_y\otimes\sigma_y)\rho^*(\tau)(\sigma_y\otimes\sigma_y)
\end{equation}
where the asterisk denotes complex conjugation in the standard basis
$\{|00\rangle,|01\rangle,|10\rangle,|11\rangle\}$ and the Pauli
matrix $\sigma_y=2I_y.$ The concurrence of the two--spin system with
 the density matrix $\rho(\tau)$ is equal to~\cite{wootters}
\begin{equation}
\label{conc} C=max\{0,
2\lambda-\lambda_1-\lambda_2-\lambda_3-\lambda_4\},\qquad
\lambda=max\{\lambda_1,\lambda_2, \lambda_3, \lambda_4\}
\end{equation}
where $\lambda_1,$ $\lambda_2,$ $\lambda_3,$ and $\lambda_4$ are the
square roots of the eigenvalues of the product
$\rho(\tau)\tilde{\rho}(\tau).$ Using
Eqs.~(\ref{rhot}),~(\ref{flip}),~(\ref{conc}) one obtains
\begin{equation}
\label{lam}
\lambda_{1,2}=\frac{\sqrt{1+\sin^2(2b\tau)\sinh^2\beta}\pm|\sin(2b\tau)|\sinh\beta}{4\cosh^2\frac{\beta}2},\qquad
\lambda_{3,4}=\frac1{4\cosh^2\frac{\beta}2}.
\end{equation}
As a result, the concurrence, $C$, is
\begin{equation}
C=\frac{|\sin(2b\tau)|\sinh\beta-1}{2\cosh^2\frac{\beta}2}.
\end{equation}
The entangled state can appear only at $\sinh\beta>1$ when the
intensity of the MQ coherence of the second order has the maximal
value. This condition means that the entanglement appears at
temperatures
\begin{equation}
\label{temp} T<\frac{\hbar\omega_0}{k\ln(1+\sqrt2)}.
\end{equation}
If one takes $\omega_0=2\pi 500\cdot10^6 s^{-1}$ the entangled state
emerges at the temperature $T_E\approx27mK.$ It is interesting to
notice that in a linear chain of dipolar coupled nuclear spins in
the thermodynamic equilibrium state, entanglement appears only at
microkelvin temperatures \cite{pyr}.

\begin{figure}[h]
\begin{center}
\includegraphics[width=8cm]{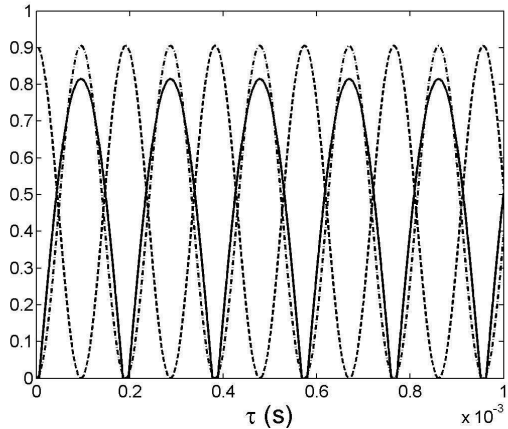}
\end{center}
\caption{The dependencies of the MQ coherences and the concurrence
on the time of the preparation period, $\tau,$ of the MQ NMR
experiment at $\beta=3.$ The coupling constant is equal to
$b=2\pi1307\, \textrm{s}^{-1};$ solid line -- concurrence; dashed
line -- intensity of the MQ coherence of the zeroth order;
dash-point line -- $G_2(\tau)+G_{-2}(\tau)$ (see the text).}
\label{graf}
\end{figure}

\begin{figure}[h]
\begin{center}
\includegraphics[width=8cm]{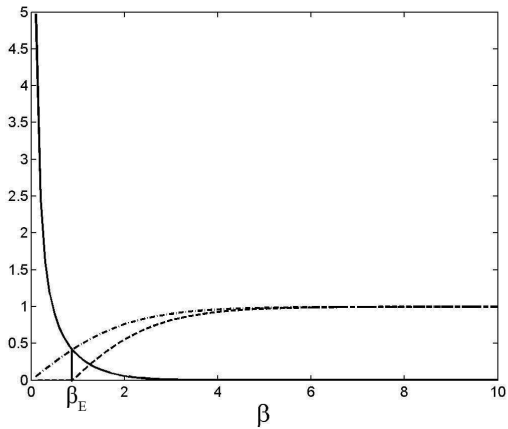}
\end{center}
\caption{The dependence of the concurrence (dashed line) and the
maximal value of $G_2(\tau)+G_{-2}(\tau)$ (dash-point line) on the
parameter $\beta$. The solid line describes the function
$[2\sinh\beta\cosh^2\frac{\beta}2]^{-1}.$ Here the coupling constant
is equal to $b=2\pi1307\,\textrm{s}^{-1}.$The entangled state
emerges at temperatures less than $T_E=\hbar\omega_0/(k\beta_E).$}
\label{grafa}
\end{figure}

The simple connection between the concurrence, $C,$ and the
intensities of the MQ coherences of the plus/minus second orders,
$G_{\pm2}(\tau),$ can be found from Eqs.~(\ref{cohs}),~(\ref{conc}):
\begin{equation}
\label{cc}
C=\sqrt{\tanh\frac{\beta}2[G_2(\tau)+G_{-2}(\tau)]}-\frac1{2\cosh^2\frac{\beta}2}.
\end{equation}
Thus, entanglement is possible only when
\begin{equation}
G_2(\tau)+G_{-2}(\tau)>\frac1{2\sinh\beta\cosh^2\frac{\beta}2},
\end{equation}
and Entanglement Witness (EW) can be introduced as the following
\begin{equation}
\label{ew}
EW=\frac1{2\sinh\beta\cosh^2\frac{\beta}2}-\{G_2(\tau)+G_{-2}(\tau)\}.
\end{equation}
In the initial moment of time $G_2(0)+G_{-2}(0)=0, EW>0$ and the
considered system is in a separable state. In the course of the MQ
NMR experiment the intensity of the MQ coherence of the second order
grows and the entanglement witness, $EW,$ changes its sign. It means
that an entangled state appears. According to \eq{cohs} the
intensities of the MQ coherences periodically change in time. The
sign of $EW$ changes also periodically. Thus separable and entangled
states change periodically in the considered system. The time
evolutions of the MQ coherences of the zeroth and second orders
together with the corresponding concurrence are represented in
Fig.~\ref{graf} at $\beta=3.$
One can see that the concurrence is close to the sum of the MQ
coherences of the plus/minus second orders,
$G_2(\tau)+G_{-2}(\tau),$ at almost all durations of the preparation
period of the MQ NMR experiment. At large $\beta$ (small
temperatures) the expression
$[2\sinh\beta\cosh^2\frac{\beta}2]^{-1}$ tends to zero and the
maximal value of $G_2(\tau)+G_{-2}(\tau)$ tends to one. This means
that the concurrence coincides with the maximal value of
$G_2(\tau)+G_{-2}(\tau)$ at small temperatures. The corresponding
dependencies of the concurrence and the maximal value of the sum
$G_2(\tau)+G_{-2}(\tau)$ on $\beta$ are given in Fig.~\ref{grafa}.
We can conclude that the entangled states appear in MQ NMR
experiments at sufficiently small temperatures. In contrast to
Ref.~\cite{souza} we study entanglement in a system of nuclear
spins(not electron ones). Such systems are more robust to
decoherence which causes the loss of the quantum information which
was obtained during quantum computation. A problem, related to ours,
was studied in Ref.~\cite{dor}. That work focuses on the high
temperature regime in which entanglement is absent. The prediction
of the existence of an entangled state at high temperatures
\cite{dor} is an artifact of an incorrect choice of the initial
density matrix.

\section{Conclusion}
MQ NMR experiments can be used for the analysis of entangled states
in spin systems. We have introduced an Entanglement Witness using
observable intensities of the MQ NMR coherences of the plus/minus
second order and analyzed entanglement in term of the Wootters
criterion. Entangled states emerge when the sum of intensities of
the MQ coherences of the plus/minus second order exceeds an exactly
calculated threshold depending on the external magnetic field and
the temperature. MQ NMR can be considered as a new method for
obtaining entangled states in spin systems.

\section{Acknowledgments}
The authors wish to express their gratitude to S.I. Doronin and M.
A. Yurishchev for many helpful discussions. This work is supported
by the Russian Foundation for Basic Research through the grant
07-07-00048.


\begin{thebibliography}{99}
\bibitem{chuang} M. A. Nielsen \and I. L. Chuang, Quantum Computation and Quantum
Information (Cambridge University Press, 2000).
\bibitem{bvsw} C. H. Bennett, D. P. DiVincenzo, J. A. Smolin, W. K.
Wootters, Phys. Rev. A {\bf 54}, 3824 (1996).
\bibitem{amico} L. Amico, R. Fazio, A. Osterloh, V. Vedral, arxiv:
quant-ph/0703044;
\bibitem{ghosh} S. Ghosh, T. F. Rosenbaum, G. Aeppll, S. N.
Coppersmith, Nature {\bf 425}, 48 (2003).
\bibitem{souza} A. M. Souza, M. S. Reis, D. O. Soares-Pinto, I. S.
Oliveira, R. S. Sarthour, Phys. Rev. B {\bf 77}, 104402 (2008).
\bibitem{horodeski} M. Horodeski, P. Horodeski, R. Horodeski, Phys.
Lett. A {\bf 223}, 1 (1996).
\bibitem{wang} X. Wang, Phys. Rev. A {\bf 66}, 034302 (2002).
\bibitem{wiesniak} M. Weisniak, V. Vedral, C. Brukner, New. J. Phys
7, 258 (2005).
\bibitem{baum} J. Baum, M. Munowitz, A. N. Garroway, A. Pines, J.
Chem. Phys. {\bf 83}, 2015 (1985).
\bibitem{lacelle} E. B. Fel'dman, S. Lacelle, J. Chem. Phys. {\bf
107}, 7067 (1997).
\bibitem{wootters} W. K. Wootters, Phys. Rev. Lett. {\bf 80}, 2245
(1998).
\bibitem{goldman} M. Goldman, Spin Temperature and Nuclear Magnetic
Resonance in Solids (Clarendon, Oxford, 1970).
\bibitem{maximov} E. B. Fel'dman, I. I. Maximov, J. Magn. Reson. {\bf
157}, 106 (2002).
\bibitem{yesinowski} G. Cho \and J. P. Yesinowski, J. Phys.
Chem. {\bf 100}, 15716 (1996).
\bibitem{pyr} Doronin S. I., Pyrkov A. N. \and Fel'dman E. B., JETP Letters {\bf
85}, 519 (2007).
\bibitem{dor} S. I. Doronin, Phys. Rev. A {\bf 68}, 052306 (2003).
\end{thebibliography}
\end{document}